\def\AaA{{\em Astr.~Astrophys.}}

\def\etal{{et al.\thinspace}}

\def\spose#1{\hbox to 0pt{#1\hss}}
\def\approxlt{\mathrel{\spose{\lower 3pt\hbox{$\sim$}}
	\raise 2.0pt\hbox{$<$}}}
\def\approxgt{\mathrel{\spose{\lower 3pt\hbox{$\sim$}}
	\raise 2.0pt\hbox{$>$}}}

\def\multleft#1{\hbox to size{\vbox {\halign {\lft{##}\cr #1}}\hfill}\par}
\def\multright#1{\hbox to size{\vbox {\halign {\rt{##}\cr #1}}\hfill}\par}
\def\Mdot{\hbox{$\dot M$}}
\def\mdot{\hbox{$\dot m$}}

\def\today{\ifcase\month\or January\or February\or March\or April\or May\or
      June\or July\or August\or September\or October\or November\or December\fi
      \space\number\day, \number\year}
\def\$<${\thinspace}
\def\s{\hbox{\phantom{5}}}	%one space
\def\boxit#1{\vbox{\hrule\hbox{\vrule\kern3pt\vbox{\kern3pt
          #1 \kern3pt}\kern3pt\vrule}\hrule}}

\def\cm{{\rm\thinspace cm}}

\def\erg{{\rm\thinspace erg}}

\def\K{{\rm\thinspace K}}
\def\keV{{\rm\thinspace keV}}
\def\km{{\rm\thinspace km}}

\def\Mpc{{\rm\thinspace Mpc}}
\def\Msun{\hbox{$\rm\thinspace M_{\odot}$}}

\def\s{{\rm\thinspace s}}

\def\sr{{\rm\thinspace sr}}

\def\ergps{\hbox{$\erg\s^{-1}\,$}}

\def\kmps{\hbox{$\km\s^{-1}\,$}}

\def\pcmsq{\hbox{$\cm^{-2}\,$}}

\def\pmpccu{\hbox{$\Mpc^{-3}\,$}}
\def\ps{\hbox{$\s^{-1}\,$}}

\def\psr{\hbox{$\sr^{-1}\,$}}

\def\kmpspMpc{\hbox{$\kmps\Mpc^{-1}$}}

\documentstyle[psfig]{mn}
\title[The X-ray Background]
{On the Origin of the Hard X-Ray Background}

\author[T. Di Matteo et al.]
{T.~Di~Matteo$^{1,2}$\thanks{AXAF Fellow}, A.~Esin$^{2,3\star}$, A.~C.~Fabian$^1$, R.~Narayan$^2$\\
{\small $^1$Institute of Astronomy, Madingley Road,
Cambridge, CB3 OHA}\\
{\small $^2$ Harvard-Smithsonian Center for Astrophysics, 60 Garden St, Cambridge, MA 02138}\\
{\small $^3$ Theoretical Astrophysics, Caltech 103-33, Pasadena, CA 91125}}

\date{}

\begin{document}

\maketitle

\begin{abstract}

Several models for the hard X-ray Background (XRB) suggest that it is 
due to the emission from heavily obscured AGN. Recent studies have
revealed the presence of a new population of hard X-ray sources which
must contribute significantly to the XRB. However, whether the
majority of these sources are obscured AGN or some other type of
object at present remains unclear.  Here, we further examine the
possibility that a significant fraction of the XRB comes from a
population of galaxies undergoing advection-dominated accretion in the
high--$\mdot$ regime and thus produce intrinsically hard spectra. When the
accretion rate is close to $\dot{m}_{\rm crit}$, above which an
advection--dominated accretion flow (ADAF) no longer exists, the major
contribution to X-ray emission is due to inverse Compton scattering of
the soft seed photons produced by cyclo--synchrotron radiation.  In
this regime, the resulting ADAF spectra are relatively hard with a
fairly constant X-ray spectral index $\alpha \sim 0.2-0.4$ and a
spectral cut-off at $\sim 200 \keV$.  We show that the integrated
emission from such sources can provide a good fit to the hard ($> 2
\keV$) X-ray background, provided that the spectrum is dominated by
the contribution from objects located at redshifts $z \sim 2-3$.  The
model requires most of the contribution to the XRB to be due to
objects accreting at $\mdot_{\rm crit}$.
\end{abstract}


\section{Introduction}

The origin of the X-ray background (XRB) is still largely an unsolved
problem.  Recent progress indicates that the XRB is probably due to
contributions from different types of discrete sources.  Although
Active galactic Nuclei (Seyferts and Quasars, hereafter AGN) provide a
large fraction of the soft XRB below $\sim$ 2 \keV ~(Shanks \etal
1991; Hasinger \etal 1998; Schmidt et al. 1998) their spectra are too
steep to explain the hard ($> 3\keV$) XRB.  Also, as a larger sample
of QSOs became available, detailed studies of the QSO X-ray luminosity
function (Boyle \etal 1994) and the source number count distribution
(Georgantopoulos et al. 1996) have shown that QSOs are unlikely to
form more than 50 per cent of the XRB, even at 1 \keV.  Models based
on the unified Seyfert scheme, in which a large fraction of the
emission from AGN is absorbed by obscuring matter (Madau \etal 1994;
Celotti \etal 1995; Comastri \etal 1995) seem to account for the
amplitude and the general shape of the hard XRB.  However, in these
models it is not clear whether cosmological evolution can ultimately
smooth out the predicted absorption edges and emission lines
(e.g. Gilli et al. 1998) in the $2-10 \keV$ spectral range, produced by
iron and other metals (Matt \& Fabian, 1994), to a level which is
consistent with the XRB spectrum statistical errors. In fact, recent
ASCA observations (e.g. Gendreau et al. 1995) showed that the spectrum
of XRB is typically very smooth and well approximated by a power law
of spectral index $\alpha=0.4$.  Thus, to explain the origin of XRB we
might have to postulate the existence of a new population of hard
X-ray emitting sources.

Recent deep ROSAT surveys have begun to resolve such sources at the
faintest soft--X--ray flux limits (Vikhlinin \etal 1995; Boyle \etal
1995; Almaini \etal 1996; Hasinger et al. 1998; 
McHardy et al. 1998). Many of these sources appear to be associated
with faint active galaxies with implied X-ray luminosities
100 times higher than normal field galaxies. Furthermore, a promising
population of sources with hard X-ray spectra has been discovered by
ASCA and BeppoSAX in the $2-10\keV$ band (Ueda et al. 1998; Giommi et
al. 1998). These findings indicate the presence of a population which
could significantly contribute to the XRB (resolving $\sim 30$ per
cent of the $2-10 \keV$ XRB). Although it seems plausible that a
fraction of the observed hard--spectrum sources might simply be
heavily obscured AGN, in the majority of cases the true nature of the
emission remains unclear. In addition, most models predict that the bulk
of the hard XRB originates at redshifts $z \sim 2$ which are currently
largely unexplored in X-rays.

Continuing on the lines of Di Matteo \& Fabian (1997; hereafter DF),
in this paper we further examine the possibility that the sources
contributing to the XRB have intrinsically hard X-ray
spectra, such as are naturally produced in the context of
advection--dominated accretion models.

In an advection-dominated accretion flow (ADAF) most of the energy
released by viscous dissipation is stored within the gas and advected
inward with the accreted plasma; only a small fraction is radiated
away (see Narayan, Mahadevan \& Quataret 1998 for a recent review).
Recent work on ADAFs has concentrated on low-$\Mdot$
solutions (Ichimaru 1977; Rees et al. 1982; Narayan \& Yi 1994, 1995;
Abramowicz et al. 1995) which exist when the accretion rate is lower
then a critical value, $\mdot_{\rm crit} \sim 1.3 \alpha^2$, in
Eddington units. This optically--thin branch is based on two critical
assumptions: (1) viscous energy is deposited into the ions and only a
small fraction of energy goes directly into the electrons; and (2) the
energy transfer between the ions and electrons occurs only via Coulomb
collisions. Because Coulomb transfer in the the low-density ADAF is
inefficient, as a result of these assumptions, the accreting gas takes
the form of a two-temperature plasma, in which the ions are at nearly
virial temperature $T_{\rm i} = 10^{12}/r\K$ and the electron
temperature saturates at around $10^9-10^{10}\K$.

The spectrum from an optically--thin ADAF is very different from that
of a standard thin disk.  The high electron and ion temperatures and
the presence of an equipartition magnetic field allow a variety of
radiation processes to contribute to the emitted spectrum from radio
to gamma rays.  The radio to X-ray emission is produced by the
electrons whereas the gamma-ray emission is produced by the ions via
pion production.  Here, we restrict the discussion to electron
emission which comprises synchrotron, inverse Compton, and
bremsstrahlung processes and in particular to those processes which
contribute to the X--ray emission.

In this {\it Letter}, we consider whether a significant fraction of
the hard XRB can be due to the integrated emission from a population
of galaxies undergoing advection--dominated accretion with $\mdot \sim
\mdot_{\rm crit}$. When the accretion rate $\dot{m}$ is substantially
lower than $\dot{m}_{\rm crit}$ the X-ray emission from an ADAF is
dominated by bremsstrahlung radiation. As the accretion rate $\dot{m}$
approaches $\dot{m}_{\rm crit}$, Comptonized emission contributes more
significantly to the the high energy spectrum.  In a previous paper DF
have analyzed the contribution to the XRB from ADAFs in the
bremsstrahlung--dominated regime. Here we show that for sources 
undergoing accretion close to $\mdot \sim \mdot_{\rm crit}$, thermal
Comptonization is highly saturated, and produces hard spectra which
can easily account for the XRB (this condition then resembles that
proposed on ad hoc basis by Zdziarski, 1988).

In \S2 of this paper, we compute the detailed spectrum of an ADAF in
the high--$\mdot$ regime using the model described by Narayan, Barret
\& McClintock (1997) and Esin, McClintock \& Narayan (1997). In
\S3, we use an appropriate cosmological model to derive the
contribution to the XRB intensity from these sources and compare the
resulting spectrum with the data. We discuss possible implications of
our model and summarize our results in \S4.  
%Good
%agreement is obtain if objects accrete within a restricted range of
%$\mdot$.

\begin{figure}
\centerline{\psfig{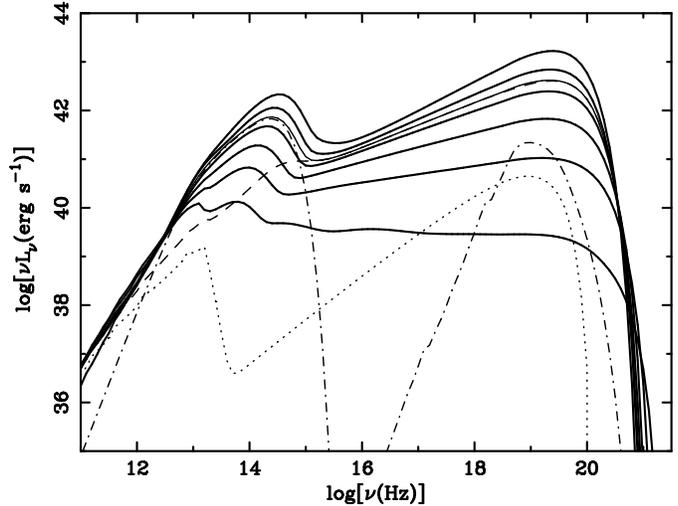}}
\caption{ADAFs accreting $\dot{m} \approxlt \dot{m}_{\rm crit}$. 
The seven solid lines show the  spectra for 
$\dot{m}=0.005,0.02$$,0.04,$$0.06,0.07,0.08,$$0.089 (=\dot{m}_{\rm crit})$, $\log(r_{\rm tr})=2.7$, $\alpha=0.25$, $\beta=0.5$.  The different spectral
components are shown for illustration only for the case of
$\dot{m}=0.07$: the synchrotron and bremsstrahlung emission -- dotted
line, the blackbody emission from the outer thin-disk and the
reflection component -- dash--dotted line, the sum of the
Comptonization of the soft disk blackbody and synchrotron photons --
dashed line. The central black hole mass is $m=10^7$.}
\label{annspec}
\end{figure}

\section{Modeling the X--ray emission from an ADAF} 
The hot, two--temperature, advection--dominated solution on which our
model is based is known to exist only for accretion rates below the
critical value $\mdot_{\rm crit}$ (e.g. Narayan \& Yi 1995; Abramowicz
et al.  1995). For $\mdot > \mdot_{\rm crit}$ Coulomb coupling becomes
very efficient, allowing the protons to cool rapidly and causing the hot
flow to collapse into a standard cooling--dominated thin disk.  In this
paper we consider a regime of accretion close to $\mdot_{\rm crit}$.

In our model, the accretion flow is composed of two separate zones: an
inner ADAF and an outer thin disk.  The boundary between the two zones
lies at the transition radius, $r_{\rm tr}$ (hereafter all radii are in
Schwarzschild units).  Since we are interested in the high $\mdot$
regime, we adopt a rather low value of the transition radius in our
calculations, setting $r_{\rm tr}=10^{2.7}$. 
It is worth noting, however, that the ADAF spectrum in the X-ray band in
practically independent of our choice of the transition radius, as long
as it lies outside $\sim 100$ Schwarzschild radii (see Esin, McClintock
\& Narayan 1997). 

We compute the ADAF spectra using the numerical procedure described in
detail by Esin, McClintock \& Narayan (1997) and Narayan, Barret \&
McClintock (1997).  This procedure combines a fairly sophisticated
treatment of radiative transfer (Poutanen \& Svensson 1996) with
global dynamical accretion flow solutions (Narayan, Kato, \& Honma
1997), while solving for the thermal balance of the ions and
electrons. In our calculations, we take the black hole mass to be $m =
10^7$ (in solar mass units), and set the viscosity parameter to
$\alpha=0.25$.  We also assume that the magnetic field in equipartition
with thermal pressure, so that the ratio between the thermal and total
pressure, denoted by $\beta$, is equal to $0.5$.

Spectra produced from such an accretion flow, for a series of $\mdot$
in the vicinity of $\mdot_{\rm crit}$, are shown in Fig.~\ref{annspec}.
The low energy (radio-IR) end of the spectrum is due to self-absorbed
synchrotron emission from hot electrons in the equipartition magnetic
field. The quasi--thermal emission from the outer thin disk peaks in
the NIR/optical band. The soft synchrotron and blackbody photons
inverse Compton scatter off the hot thermal electrons in the ADAF and
produce the hard power law spectrum with a spectral cut-off at $\sim
2-3 kT_{\rm e}$. X-rays scattering off the outer thin disk produce a
reflection bump at a characteristic energy $\sim 30\ {\rm keV}$.  We
also include bremsstrahlung emission which peaks at $h\nu
\approx kT_{\rm e}$. 

The relative importance of Comptonization increases very rapidly with
increasing $\mdot$ (see Fig.\ref{annspec}).  In the high $\mdot$ regime
considered here, the characteristic electron scattering optical depth
$\tau$ of the ADAF becomes of order unity, since $\tau \propto \mdot$.
This implies that the value of the Compton $y-$parameter
%, which can be
%written $y = (\tau+\tau^2) (4 \theta_{\rm e} + 16 \theta_{\rm e}^2$
%where $\theta_{\rm e} = k T_{\rm e}/(m_e c^2)$ is the dimensionless
%electron temperature (Rybicki \& Lightman 1979), 
is close to unity as
well.  As a consequence, inverse Compton scattering becomes the most
important cooling mechanism, and the Comptonized tail dominates the
X-ray spectrum, as shown by the models plotted in Fig.~\ref{annspec}.

The energy boost experienced by a photon in each scattering is
proportional to the electron temperature. Although, with increasing
$\mdot$ more energy is transferred to the electrons by Coulomb
coupling, inverse Compton cooling also becomes more efficient, with
the result that $T_{\rm e}$ saturates at around a few times $10^9 \K$.
Since the synchrotron radiation (the main source of seed photons) is
very soft (see Fig.~\ref{annspec}), the photons will be scattered many
times before reaching the Wien saturation limit, giving rise to a
smooth power law spectrum. The slope ($\alpha_{x}$) of the power law
is proportional to $\log{\tau}$, so that the X-ray spectrum becomes
flatter (in $L_{\nu}$) as $\mdot$ increases.

The hard X-ray spectrum of an ADAF closely resembles that of the
hard X--ray background when the accretion rate is in the vicinity of
$\mdot_{\rm crit}$, but the spectra from sources with lower $\mdot$
become considerably steeper (as illustrated in Fig.~\ref{annspec}).
However, because the inverse Compton cooling increases as $\mdot^2$
(see Fig.~\ref{annspec}), eventual contributions to the XRB intensity
from systems with softer spectra (i.e. smaller $\mdot$) is not very
important, once we assume that a population of sources accreting at
$\mdot=\mdot_{\rm crit}$ exists.  Since the sources with flat spectra,
accreting close to $\mdot_{\rm crit}$, are more luminous, they
contribute more significantly to the XRB intensity.  To make this
statement more quantitative, the contribution from the lower $\mdot$
regime is negligible, unless the relative fraction of the
steep--spectrum sources is greater than that for the hard--spectrum
sources by a factor comparable do the difference in their respective
hard X-ray luminosities.  The latter condition is satisfied only is we
draw our sources from a population with number density per logarythmic
interval in $\mdot$ described by $N(\mdot) \propto \mdot^{-i}$, where
$i \ge 2$.
   
In the next section we will consider the integrated emission from
sources producing hard Comptonized spectra in the high $\mdot$ regime
consistent with a two temperature ADAF, and fold it with the
appropriate cosmological model to constrain their potential
contribution to the unresolved $3-100 \keV$ XRB.

\section{Contribution to the XRB}
\subsection{Cosmological Model}
In order to obtain a model for the XRB, we consider the contribution
from many unresolved sources accreting at $\mdot \sim \mdot_{\rm
crit}$, distributed over a redshift range from the local universe
($z=z_0$) to the early universe ($z=z_{\rm max}$).  Comoving spectral
emissivity from a distribution of such objects can be written as a
product $j[E,z]=n(z)L_{\rm E}(z)$, where $n(z)$ is the
comoving number density of X-ray sources, and $L_{\rm E}(z)$ represents
the specific luminosity of individual sources (see \S2).  In general,
both $n$ and $L_{\rm E}$ are functions of redshift, so that neither
pure luminosity nor pure number density evolution are very good
assumptions.  Instead, we adopt the following simple prescription for
the redshift evolution of comoving emissivity, $j[E,z]=j_0 (E)
(1+z)^{p}$, where $j_0 (E)$ is the local (at $z=z_0$) spectral
emissivity and $p$ is the evolution parameter.  

The total flux received from such objects, calculated in the
framework of the Robertson-Walker metric, is given by
\begin{equation}
\label{xrb_mod}
I(E)=\frac{c}{4\pi H_0}\times \int_{z_0}^{z_{\rm
max}}\frac{(1+z)^{p-2}}{(1+2q_0z)^{1/2}}j_0[E(1+z)]dz,\nonumber
\end{equation}   
where $q_0$ is the deceleration parameter, $H_0$ the Hubble constant
(we use $q_0=0.5$ and $H_0=65 \kmpspMpc$).  $I(E)$ represents our
computed XRB intensity (in units of $\keV\ps\psr\pcmsq\keV^{-1}$).
The only free parameter in Eqn. (\ref{xrb_mod}) is the local source 
number density, $n(z_0)$, which we determine by 
comparing $I(E)$ with the empirical fit to the XRB spectrum.

\begin{figure}
\centerline{\psfig{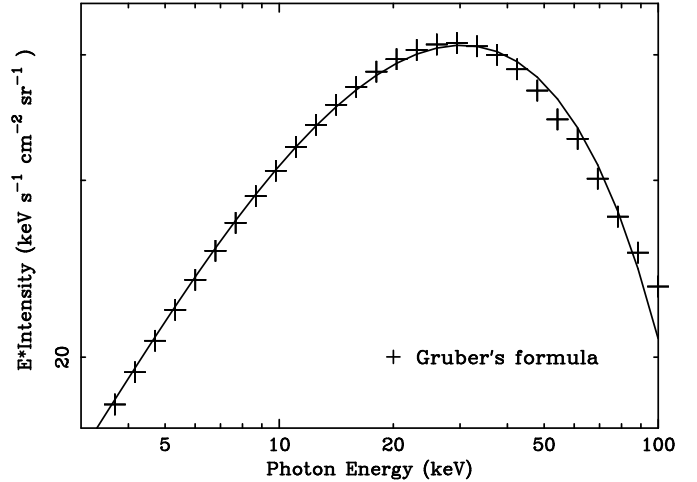}}
\caption{The spectrum of the X-ray background (Gruber's empirical formula 
indicated by crosses) compared to our best fit model spectrum
calculated with advection-dominated flows accreting at $\mdot \sim
\mdot_{\rm crit}$ and cosmological evolution parameters fixed at
$z_{\rm max}=3$, $z_0=2$, and $p=0$.}

%{The X-ray background from advection-dominated  flows 
%accreting at $\dot{m} \sim \dot{m}_{\rm crit}$, compared to Gruber's
%(1992) empirical formula (Eqn. \ref{xrb_mod}, \ref{xrb_obs}) from 3 to
%100 \keV. Six models are plotted. 
%%The heavier (solid and dashed)
%%lines only consider the ADAF model sources; the thinner (dotted and
%%dash-dotted) lines take into account an additional contribution of
%%about 20 per cent (in the $2-10 \keV$ band) from a power law component
%%of energy index $\alpha = 0.7$ due to AGN. 
%The dashed and dotted
%lines take into account accretion at different $\mdot = 0.005, 0.02,
%0.04, 0.6,0.7,0.8,\mdot_{\rm crit}$ (as shown in Fig.~\ref{annspec})
%in each $z$ shell volume.  The solid and dot-dashed lines only
%consider sources accreting at $\mdot =\mdot_{\rm crit}$.  The thinner
%solid and dashed lines for are for $z_{\rm max}=3.5$, $z_0=0$ and
%$p=3$. The heavier lines  are for $z_{\rm
%max}=3$, $z_0=2$ and $p=0$.  The overall fit from all of these models
%is quite good in particular in the $5-100 \keV$ band. The goodness of
%fit from the different models can be more directly assessed from the
%residual plot in Fig.~\ref{resi}.}

\label{highmdotfit}
\end{figure}

\begin{figure}
\centerline{\psfig{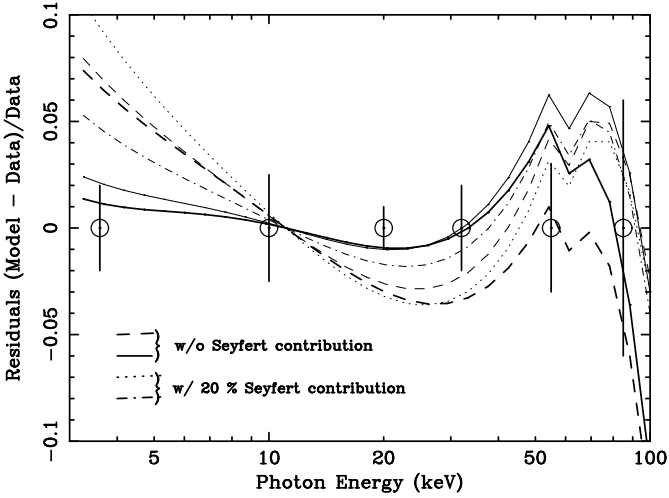}}
\caption
{Deviation of our models from the analytical approximation by Gruber
(1992). The error bars are a sample of the typical one sigma
measurement errors at various energies. Our best fit model (heavy
solid line also shown in Fig.~\ref{highmdotfit}) gives an excellent
fit to the observed XRB spectrum over most energy bands.  It is
calculated by considering only the sources with the hardest spectra
($\mdot= \mdot_{\rm crit}$) and by setting $z_{\rm max}=3$, $z_0=2$,
$p=0$.  Adding a contribution from objects accreting at different
$\mdot$ in the range between 0.005 and $\mdot_{\rm crit}$ (as shown in
Fig.~\ref{annspec}; dotted and dashed lines) steepens the
overall spectrum and the fit worsens at the lower energies ($<5
\keV$).  Adding a 20 per contribution to the XRB from Seyfert galaxies
with a canonical spectral slope of $\alpha=0.7$ (shown as a dot-dashed
line for $\mdot=\mdot_{\rm crit}$ model and as a dotted line for the
model with a range of $\mdot$) further increases the excess at soft
energies.  To illustrate the effects of cosmology, we also plot
(thinner solid and dashed line) the spectrum computed for the same
model as that shown by the heavy solid line, but with different
cosmological evolution parameters: $z_{\rm max}=3.5$, $z_0=0$, $p=3$.}

%{Percentage deviations of our best fit models from 
%Fig.~\ref{highmdotfit} from the analytical approximation by Gruber,
%1992. The error bars are a sample of the typical one sigma measurement
%errors at various energies. 
%Our best fit models (thin and heavy solid
%lines) give an excellent fit to the observed XRB spectrum over most
%energy bands. For these models we only consider the hardest, $\mdot=
%\mdot_{\rm crit}$, source spectra. Adding a contribution from
%objects accreting at different $\mdot$ (thin and heavy dashed lines)
%steepens overall spectrum and the fit worsens at the lower energies
%($<5 \keV$). A 20 per cent contribution to the XRB due to Seyfert
%galaxies with a canonical spectral slope of $\alpha=0.7$ (dot-dashed
%and dotted lines, for respective ranges of $\mdot$ distribution; see
%caption of Fig.~\ref{highmdotfit}) added to our model spectra further
%increases the excess at soft energies.}
\label{resi}
\end{figure}

\subsection{Comparison with the data}
An empirical fit to the spectrum above $\sim 3 \keV$ where most of the
energy density resides, has been determined by Gruber (1992). The
energy flux in units of $\keV \ps\psr\pcmsq\keV^{-1}$ is given by:
\begin{equation} 
\label{xrb_obs}
I_{\rm obs}(E) = \left\{ \begin{array}{ll} 7.877E^{-0.29}{\rm exp}
(-E/41.13\keV), \\ 
             \hspace{0.8in} 3\keV < E < 60\keV; \vspace{0.2cm} \\ 
        1652 E^{-2.0} + 1.754 E^{-0.7}, \\ 
             \hspace{0.8in} 60\keV < E < 100\keV. \end{array} \right.
\end{equation}
Fig.~\ref{highmdotfit} shows a comparison between the data and our
best fit model for the predicted contribution of the ADAFs to the
background.  Fig.~\ref{resi} shows the fractional difference between
our model results and the analytical approximation to the XRB in the
energy range $3-100\keV$ (Eqn. \ref{xrb_obs}). To judge the goodness
of fit,we show a few typical measurements errors at different
energies. The different fits (described in detail below), which can be
identified from the plot of their residuals in Fig.~\ref{resi},
correspond to essentially two different models.

In one model, corresponding to the dashed and dotted line in
Fig.~\ref{resi}, we take into account
contributions from sources accreting at a range of mass accretion
rates (specifically, we consider sources whose spectra are shown in
Fig. \ref{annspec} contributing in equal proportion) in each
successive redshift shells.  As we argued in
\S2, the contribution from objects with $\mdot$ lower than the values
considered here is negligible as long as the relative fraction of
sources at a given $\mdot \approxlt \mdot_{\rm crit}$ does not exceed
the number of sources close to $\mdot_{\rm crit}$ by a factor
corresponding to $\sim L_{\mdot_{\rm crit}}/L_{\mdot}$. The thinner
dashed-line corresponds to a model with $z_{\rm max}=3.5$ and $z_0=0$.
In this case, the normalization of the XRB spectrum requires a local
comoving number density of $n(z_0=0) \sim 10^{-4}\pmpccu $
%(roughly
%comparable with the present day number of Seyfert 1 and 2
%galaxies). 
The best fit evolutionary parameter is $p=3$ (similar to
that inferred for AGN), implying that most of the contribution to the
XRB intensity comes from objects at high redshifts. To illustrate this
further, we show a model (thicker dashed line) in which all of the emission
comes from $z_{\rm max}=3$ and $z_0=2$ and the evolution parameter $p$
is set to be equal to 0. In this case, the comoving number density of
sources, at these redshifts, is consistently higher and of the order of
$n(z_0=2) \sim 10^{-2}\pmpccu $. The quality of the two fits (one with
strong evolution and the other with a single population of sources
between $z=2$ and $3$) is very similar. For $E
\approxgt 5 \keV$ the computed intensity is within $2 \sigma$ of the
data points, at $E
\approxlt 5\keV$, it significantly overestimates the observed
spectrum.  However, small amounts of absorption (i.e. a column
density $N_{\rm H} \approxlt$ a few $10^{21} \pcmsq$ -- much less than that
required by Seyfert models of the XRB) would significantly improve the
fit at these low energies, where the emission from other extragalactic
sources also becomes relevant.

If we only consider the emission from the brightest sources, those
with $\mdot \sim \mdot_{\rm crit}$ we find that the fit to the low
energy range of the XRB significantly improves: the computed intensity
now lies within $1 \sigma$ of the data points (solid lines in
Figs.~\ref{highmdotfit} and ~\ref{resi}). The spectra from ADAF
sources accreting at these high $\mdot$ values are very hard and give
an excellent fit to the XRB when appropriately redshifted.  Again, we
compute the model intensity considering first, a strongly evolving
population ($p=3$) from $z_{\max} = 3.5 $ to $z_0=0$ (thinner solid
line) and second, a short lived, non-evolving ($p=0$), population of
objects from $z=3$ to $z_0=2$ (heavier solid line; this our best fit
model  plotted in Figs.~\ref{highmdotfit}). The normalizations to
the XRB yield comoving number densities very similar to those obtained
above.

The same set of models described above (for $z_{\rm max}=3.5, z_{0}=0$
and $p=3$), is also calculated considering a 20 per cent contribution
to the $2-10 \keV$ XRB due to AGN characterized by a canonical
power-law spectrum with energy index $\alpha_{\rm x}=0.7$. Such a
contribution produces a further steepening of the spectrum and
therefore a greater excess at soft energies ($3-5 \keV$ range; dotted
and dashed line in Fig.~\ref{resi}).  However, the absorption column
density required to suppress such an excess is still much smaller than
that required by Seyfert models of the XRB and of the order of $N_{\rm
H} \approxlt 10^{22}\pcmsq$.  Note that because of the outstanding
issue that the observed AGN spectrum does not match that of the XRB
(there is no strong evidence for any hardening of the X-ray spectral
slope even for AGN identified in deep ASCA fields; Boyle et al. 1998),
we know that there has to be a very significant residual contribution
to the $2-10
\keV$ XRB from sources that we have yet to identify (which we suggest
might be ADAFs).

In summary, we have directly shown that a significant contribution to
the XRB can result from a population of galaxies undergoing
advection dominated accretion in their nuclei with $\mdot \sim
\mdot_{\rm crit}$. In this high--$\mdot$ regime an ADAF is luminous
enough to satisfy the energy constraints of the XRB (in the $2-10
\keV$ band the ADAF luminosities are in the range of $2-8 \times
10^{42} \ergps$ for $0.07 \approxlt \mdot \approxlt
\mdot_{\rm crit}$, for $m=10^7$ as shown 
in Fig.~\ref{annspec}); moreover, an ADAF produces a very hard
spectrum, through photon-starved Comptonization, which can explain the
observed spectrum of XRB.

\section{Conclusion and Discussion}
The main results obtained in this paper can be summarized as follows:

(a) In the framework of the ADAF paradigm it is
possible to construct a model which reproduces with good accuracy the
XRB spectrum in the range $\sim 3-100 \keV$ and meets all the
presently known constraints on number density of source
population and X-ray spectral characteristics.   

(b) The key feature of the
model is the existence of a population of ADAFs accreting at $\mdot
\sim \mdot_{\rm crit}$. In this regime thermal Comptonization of
cyclo-synchrotron photons gives rise to a typically hard X-spectrum
(with $\alpha_{x} \sim 0.2-0.4$ ) which can naturally fit the $3-100
\keV$ XRB. Also, because of the small solid angle subtended by the
cold material, any spectral features due to X-ray irradiation are
negligible, and the model is consistent with the observed smoothness
of the XRB spectrum.

(c) We compute in detail the typical luminosity and spectra for such
sources. The contribution from the Comptonized emission from a
population of ADAFs is integrated over redshift to reproduce the XRB
intensity .  Good fits are obtained for $z_{\rm max} \sim 3.5$ and
values for the evolution parameter $p \sim 3$.  Implying that the peak
of the XRB production is around $z \sim 3 - 2$ (similar quality of
fits are in-fact obtained by considering a single population of ADAFs
between $z_{\rm max}=3$ and $z_{0} = 2$ and assuming $p=0$) The
normalization to the XRB intensity yields a comoving number density of
$n_0\sim 10^{-4}\pmpccu$, similar to that of present day Seyfert
galaxies.

The requirement that $z_{\rm max} \sim 3$ suggests that the XRB is
produced before the main quasar phase, which peaks at redshift $z \sim
2$. If a population of massive black holes was assembled very early,
by $z \sim 3$, maybe in dwarf galaxies (e.g.  M32 contains a black
hole of $M_{\rm BH}= 2 \times 10^{6} \Msun$; Van der Marel 1998;
Magorrian et al. 1998) it could have undergone advection--dominated
accretion between $z \sim 3$ and 2.  It is possible that during that
time the infall of gas into the central black hole was increasing,
e.g. due to galaxy mergers or to star formation, which allowed the
sources to accrete close to $\mdot_{\rm crit}$.  In this scenario, the
end of the ADAF phase would simply be due to the transition to the
active QSO phase with $\mdot >
\mdot_{\rm crit}$.  We further speculate that once the accretion fuel
was depleted, quasars rapidly dropped again into the low-$\mdot$ ADAF
state.  In this subsequent ADAF phase, $\mdot$ is much lower and
bremsstrahlung emission can provide a further energy contribution to
the hard XRB (as discussed by DF where $z_{\rm max}$ was determined to
be $\sim 2$).

In this paper we have assumed that all systems have black holes with a
constant mass, $m = 10^7$.  Since the ADAF spectra are rather
insensitive to $m$ while X-ray luminosity changes linearly with this
parameter, adopting a varying $m$ prescription will simply give a
different value for the comoving number density of X-ray sources and
will not change any of the other results in this paper.

We must note that decreasing electron temperatures in the accretion
flow by a factor of $\sim 2-4$ would cause the spectra to have lower
cutoff energies, moving the redshift at which the bulk of XRB is
produced closer to the QSO death-line at $z \sim 2$.  We believe that
the present ADAF model is robust enough to make such a decrease in the
gas temperature unlikely.  However, some physical mechanism currently
not included in the model, for example the presence of strong winds
from the accretion flow (as suggested by Blandford \& Begelman 1998),
might produce such an effect (e.g. Di Matteo et al. 1998).

Deep X-ray observations with {\it AXAF} and later missions should
resolve objects at redshifts beyond the peak of the XRB production,
thereby determining the origin and the nature of the hard sources
contributing to XRB emission and providing clues for understanding the
evolution of the fueling of massive black holes.  If ADAFs do
contribute significantly, the absence of a strong UV bump in the
spectrum means that the objects will not have a normal broad--line
region. They will not resemble a standard, broad--line AGN in the
optical/UV band.

The main limitation of the high-$\mdot$ ADAF model (as it has become
apparent in \S3), is that it requires most of the objects to be
accreting close to $\mdot_{\rm crit}$. Although this might seem a
peculiar feature it is worth noticing that models for the galactic
black hole candidate Cyg X-1 (Esin et al. 1998), can also be explained
by an ADAF which spends most of its time accreting at $\mdot_{\rm
crit}$; this might indicate that mechanisms might exist which lock the
accretion rate in some systems at values close to their critical rate.

\section*{Acknowledgements}
TDM acknowledges partial support from PPARC and Trinity College,
Cambridge. AE was partially supported by National Science Foundation
Graduate Research Fellowship. Both TDM and AE were supported by NASA
through AXAF Fellowship grant number PF9-10005 and PF8-10002
respectively, which is operated by the Smithsonian Astrophysical
Observatory for NASA under contract NAS8-39073.
ACF thanks the Royal Society for support.

\end{document}